\documentclass[twocolumn]{aastex62}

\pdfoutput=1 
\usepackage{amsmath,amstext,psfig,color}
\usepackage{apjfonts} 

\def\ra#1#2#3{#1$^{\rm h}$ #2$^{\rm m}$ #3$^{\rm s}$}
\def\dec#1#2#3{$#1^\circ #2' #3''$}

\def\gw{GW170817}
\def\ngc{NGC4993}

\shorttitle{GW170817 HST}
\shortauthors{Fong et al.}

\begin{document}

\title{The Optical Afterglow of GW170817: An Off-axis Structured Jet and Deep Constraints on a Globular Cluster Origin}

\newcommand{\NU}{\affiliation{Center for Interdisciplinary Exploration and Research in Astrophysics and Department of Physics and Astronomy, Northwestern University, 2145 Sheridan Road, Evanston, IL 60208-3112, USA}}
\newcommand{\CfA}{\affiliation{Center for Astrophysics-Harvard \& Smithsonian, 60 Garden Street, Cambridge, MA 02138-1516, USA}}
\newcommand{\OU}{\affiliation{Astrophysical Institute, Department of Physics and Astronomy, 251B Clippinger Lab, Ohio University, Athens, OH 45701, USA}}
\newcommand{\Einstein}{\altaffiliation{NASA Einstein Fellow}}
\newcommand{\Hubble}{\altaffiliation{NASA Hubble Fellow}}
\newcommand{\Columbia}{\affiliation{Department of Physics and Columbia Astrophysics Laboratory, Columbia University, New York, NY 10027, USA}}
\newcommand{\Ferrara}{\affiliation{Dipartimento di Fisica, Universit\`a Ferrara, Via Paradiso 12, I-44100 Ferrara, Italy }}
\newcommand{\NYU}{\affiliation{Center for Cosmology and Particle Physics, New York University, 726 Broadway
New York, NY 10003, USA}}
\newcommand{\PRIC}{\affiliation{TO FILL IN-3}}
\newcommand{\Carnegie}{\affiliation{Observatories of the Carnegie Institute for Science, 813 Santa Barbara Street, Pasadena, CA 91101-1232, USA}}
\newcommand{\MSU}{\affiliation{Center for Data Intensive and Time Domain Astronomy, Department of Physics  and Astronomy, Michigan  State University, East Lansing, MI 48824}}
\newcommand{\UA}{\affiliation{Steward Observatory, University of Arizona, 933 North Cherry Avenue, Tucson, AZ 85721-0065, USA}}
\newcommand{\AAS}{\affiliation{American Astronomical Society, 1667 K~Street NW, Suite 800, Washington, DC 20006-1681, USA}}
\newcommand{\Bath}{\affiliation{Department of Physics, University of Bath, Claverton Down, Bath, BA2 7AY, UK}}
\newcommand{\Purdue}{\affiliation{Department of Physics and Astronomy, Purdue University, 525 Northwestern Avenue, West Lafayette, IN 47907, USA}}
\newcommand{\Southampton}{\affiliation{Mathematical Sciences and STAG Research Centre, University of Southampton, Southampton SO17 1BJ, United Kingdom}}
\newcommand{\IfaEdinburgh}{\affiliation{Institute for Astronomy, University of Edinburgh, Royal Observatory, Blackford Hill, EH9 3HJ, UK}}
\newcommand{\Birmingham}{\affiliation{Birmingham Institute for Gravitational Wave Astronomy and School of Physics and Astronomy, University of Birmingham, Birmingham B15 2TT, UK}}
\newcommand{\UCB}{\affiliation{Department of Astronomy and Theoretical Astrophysics Center, University of California Berkeley, Berkeley, CA 94720}}
\author[0000-0002-7374-935X]{W.~Fong}
\NU

\author{P.~K.~Blanchard}
\NU \CfA

\author{K.~D.~Alexander}
\Einstein \NU 

\author[0000-0002-1468-9668]{J.~Strader}
\MSU

\author{R.~Margutti}
\NU

\author{A.~Hajela}
\NU

\author{V.~A.~Villar}
\CfA

\author{Y.~Wu}
\NYU

\author[0000-0001-9582-881X]{C.~S.~Ye}
\NU

\author{E.~Berger}
\CfA

\author[0000-0002-7706-5668]{R.~Chornock}
\OU

\author{D.~Coppejans}
\NU

\author[0000-0002-2478-6939]{P.~S.~Cowperthwaite}
\Hubble \Carnegie

\author{T.~Eftekhari}
\CfA

\author[0000-0003-1503-2446]{D.~Giannios}
\Purdue

\author{C.~Guidorzi}
\Ferrara

\author[0000-0002-8560-692X]{A.~Kathirgamaraju}
\Purdue \UCB

\author{T.~Laskar}
\Bath

\author[0000-0002-0106-9013]{A.~MacFadyen}
\NYU

\author[0000-0002-4670-7509]{B.~D.~Metzger}
\Columbia

\author{M.~Nicholl}
\IfaEdinburgh
\Birmingham

\author[0000-0001-8340-3486]{K.~Paterson}
\NU

\author{G.~Terreran}
\NU

\author{D.~J.~Sand}
\UA

\author{L.~Sironi}
\Columbia

\author{P.~K.~G.~Williams}
\CfA\AAS

\author{X.~Xie}
\Southampton

\author{J.~Zrake}
\Columbia

\begin{abstract}
We present a revised and complete optical afterglow light curve of the binary neutron star merger \gw, enabled by deep {\it Hubble Space Telescope} ({\it HST}) F606W observations at $\approx\!584$~days post-merger, which provide a robust optical template. The light curve spans $\approx 110-362$~days, and is fully consistent with emission from a relativistic structured jet viewed off-axis, as previously indicated by radio and X-ray data. Combined with contemporaneous radio and X-ray observations, we find no spectral evolution, with a weighted average spectral index of $\langle \beta \rangle = -0.583 \pm 0.013$, demonstrating that no synchrotron break frequencies evolve between the radio and X-ray bands over these timescales. We find that an extrapolation of the post-peak temporal slope of \gw\ to the luminosities of cosmological short GRBs matches their observed jet break times, suggesting that their explosion properties are similar, and that the primary difference in \gw\ is viewing angle. Additionally, we place a deep limit on the luminosity and mass of an underlying globular cluster of $L \lesssim 6.7 \times 10^{3}\,L_{\odot}$, or $M \lesssim 1.3 \times 10^{4}\,M_{\odot}$, at least 4 standard deviations below the peak of the globular cluster mass function of the host galaxy, NGC4993. This limit provides a direct and strong constraint that \gw\ did not form and merge in a globular cluster. As highlighted here, {\it HST} (and soon {\it JWST}) enables critical observations of the optical emission from neutron star merger jets and outflows. 
\end{abstract}


\section{Introduction}

The discovery of optical light from the first binary neutron star merger, \gw\ \citep{gw170817,Arcavi17_2,cfk+17,Lipunov17,sha+17,Tanvir17,vsy+17} localized the event to a projected distance of $\approx 2$~kpc from its host galaxy NGC4993 \citep{bbf+17,llt+17}, and provided a precise position for follow-up observations across the electromagnetic spectrum (e.g., \citealt{gw170817mma}). Energy released from the radioactive decay of heavy elements synthesized in the merger ejecta (resulting in a ``kilonova''; \citealt{mmd+10}) dominated the optical emission at early times, and its characterization was primarily led by ground-based observations \citep{Andreoni17,Arcavi17_2,cbv+17,cbk+17,cfk+17,Diaz17,Drout17,Kasliwal17,Lipunov17,nbk+17,Pian17,Pozanenko17,scj+17,Tanvir17,Troja17,Utsumi17,vsy+17,vgb+17}. At a few weeks post-merger, the field became inaccessible to optical facilities. When \gw\ emerged from solar conjunction at $\approx\!100$~days, the non-thermal afterglow emission, which results from relativistic material interacting with the surrounding medium, outshined the kilonova. The study of this second phase in the optical band was enabled by the {\it Hubble Space Telescope} ({\it HST}) \citep{amb+18,lll+18,max+18,lll+19,ptz+19}, which was the only facility with the sensitivity to securely detect the source at these epochs, due to a combination of intrinsic faintness of the afterglow and contaminating light from NGC4993.

In conjunction with ongoing radio and X-ray campaigns, the optical afterglow probes the relativistic outflow from the merger. Several studies based primarily on the radio and X-ray observations of \gw\ have converged on a structured jet model, in which the bulk of the energy is carried by a relativistic jet, surrounded by less-collimated, slower material \citep{amb+18,lpm+18,max+18,mdg+18,mfd+18,tpr+18,wm18}. These studies have also shown that at $\gtrsim100$~days post-merger, the broad-band spectral energy distribution (SED) follows a single power law characterized by $F_{\nu} \propto \nu^{-0.6}$. Notably, the optical band provides an important anchor between the nine orders of magnitude in frequency from the radio to the X-ray bands.

Thus far, extracting the flux from the optical counterpart of \gw\ has relied upon modeling the surface brightness profile of NGC4993 and subtracting its contribution. However, the morphology of NGC4993 is complex, and characterized by dust lanes and concentric shells \citep{bbf+17,llth+17,pht+17}, making accurate and uniform photometry extremely challenging. Thus, previous studies which utilized optical data suffer from a combination of imperfect galaxy subtraction and non-uniform photometric methods \citep{amb+18,lll+18,max+18,lll+19,ptz+19}.
 
Here, we present a deep {\it HST}/F606W observation of \gw\ at $\approx 584$~days, which serves as the first robust optical template for the late-stage afterglow emission, against which we can subtract earlier epochs. This enables reliable and uniform photometry of the optical afterglow for the first time. We use the observation to produce a complete and revised light curve of the optical afterglow in the F606W filter, as well as a direct and strong limit on an underlying globular cluster (GC) to constrain the formation of its progenitor. In the following sections, we present the new and archival observations used in this study (\S\ref{sec:obs}), the details of the image subtraction and broad-band spectral fitting (\S\ref{sec:analysis}), a discussion of the afterglow properties in the context of short GRBs and the limit with respect to the globular cluster mass function of NGC4993 (\S\ref{sec:disc}), and concluding remarks (\S\ref{sec:conc}).

All magnitudes in this paper are in the AB system and corrected for a Galactic extinction of ($E(B-V)=0.109$; \citealt{sf11}). Reported uncertainties correspond to $68\%$ confidence. We employ a standard $\Lambda$CDM cosmology with $\Omega_M=0.286$, $\Omega_\Lambda=0.714$, and $H_0=69.6$ km s$^{-1}$ Mpc$^{-1}$ \citep{blw+14}. We adopt a distance to both \ngc\ and the afterglow of $D_L=40.7$~Mpc \citep{cjb+18}.

\begin{deluxetable*}{lcccccc}[!t]
\tabletypesize{\normalsize}
\tablecolumns{7}
\tablewidth{0pc}
\tablecaption{HST/F606W Afterglow Photometry of GW170817
\label{tab:obs}}
\tablehead{
\colhead {Mid-time}	 &
\colhead {$\delta t$}	 &
\colhead {Instrument}  &		
\colhead {Exp. Time}		    &
\colhead {AB Mag}		    &
\colhead {$F_{\nu}$}		    &
\colhead {Program ID} \\
\colhead {(UT)}	&
\colhead {(d)}	 &
\colhead {} 		&
\colhead {(s)}	 &
\colhead {}	&
\colhead {($\mu$Jy)}	&
\colhead {}	
}
\startdata
2017 Dec 6.022 & 110.49 & WFC3/UVIS2 & 2264 & $26.31 \pm 0.19$ & $0.110 \pm 0.019$ & 14270 \\
2018 Jan 1.573 & 137.04 & ACS/WFC & 2120 & $26.59 \pm 0.23$ & $0.084 \pm 0.018$ & 15329 \\
2018 Jan 29.721 & 165.19 & WFC3/UVIS & 2372 & $26.50 \pm 0.19$ & $0.091 \pm 0.016$ &14607 \\
2018 Feb 5.740 & 172.21 & WFC3/UVIS2 & 2400 & $26.58 \pm 0.22$ & $0.085 \pm 0.017$ & 14771 \\
2018 Mar 14.626 & 209.10 & WFC3/UVIS & 2432 & $26.61 \pm 0.26$ & $0.082 \pm 0.020$ &14607 \\
2018 Mar 23.895 & 218.37 & ACS/WFC & 2120 & $26.90 \pm 0.31$ & $0.063 \pm 0.018$ & 15329 \\
2018 Jun 10.327 & 296.80& WFC3/UVIS2 & 5220 & $27.29 \pm 0.35$ & $0.044 \pm 0.014$ & 14771 \\
2018 Jul 11.752$^{a}$ & 328.22 & WFC3/UVIS2 & 14070 & $27.58 \pm 0.35$ & $0.034 \pm 0.011$ & 15482 \\
2018 Jul 20.357 & 336.83 & ACS/WFC & 2120 & $\gtrsim 27.2$ & $\lesssim 0.048$  &15329 \\
2018 Aug 14.852$^{b}$ & 362.32 & WFC3/UVIS2 & 14070 & $27.83 \pm 0.29$ & $0.027 \pm 0.0072$ & 15482 \\
2019 Mar 24.659$^{c}$ & 584.13 & ACS/WFC & 26912 & $\gtrsim 28.2$ & $\lesssim 0.019$ & 15606 
\enddata
\tablecomments{Times are quoted in the observer frame. All observations are taken with the F606W filter. Limits corrrespond to $3\sigma$ confidence and uncertainties correspond to $1\sigma$. Magnitudes are corrected for Galactic extinction \citep{sf11}. \\
$^{a}$ Two separate visits on 2018 Jul 10 and 2018 Jul 13 UT. \\
$^{b}$ Two separate visits on 2018 Aug 14 and 2018 Aug 15 UT. \\
$^{c}$ Two separate visits on 2019 Mar 21 and 2019 Mar 27 UT. }
\end{deluxetable*} 

\section{Observations}
\label{sec:obs}

\subsection{A Deep F606W Observation}
\label{sec:temp}

We obtained {\it HST} observations of \gw\ with the Advanced Camera for Surveys (ACS) under Program 15606 (PIs: Fong, Margutti). The observations were performed in the F606W filter over two visits on 2019 Mar 21 UT and 2019 Mar 27 UT for a total on-source time of 26912~sec (6 orbits).

We retrieve calibrated {\tt FLC} images from the Mikulski Archive for Space Telescopes (MAST) archive\footnote{https://archive.stsci.edu/hst/; \dataset[https://doi.org/10.17909/t9-6qez-fw41]{https://doi.org/10.17909/t9-6qez-fw41}}, pre-corrected for charge transfer efficiency. We used tasks as part of the Drizzlepac software package \citep{gon12} and IRAF \citep{iraf1,iraf2} to process the data. We used the {\tt astrodrizzle} task to create a combined drizzled image for each visit, using {\tt final\_scale} = $0.05''$ pixel$^{-1}$ and {\tt final\_pixfrac} = 0.8, and then aligned the images to a common early epoch using the {\tt tweakreg} task (described in \S\ref{sec:astrometry}) with relative astrometric uncertainties of $\approx 6.5-8.0$~mas ($\approx 0.1-0.2$ {\it HST} pixels). We used IRAF/{\tt imcombine} to combine the images from both visits. The mid-time of the final combined image corresponds to $\delta t \approx 584.1$~days, where $\delta t$ is the time since the gravitational wave trigger (2017 August 17 at 12:41:04 UT; \citealt{gw170817}).
 
\subsection{Archival Observations}
Since our study concentrates on the optical afterglow emission of \gw, we retrieve images from MAST taken with ACS and the Wide Field Camera~3  (WFC3) in the F606W filter at $\gtrsim\!100$~days. The observations comprise ten epochs spanning 2017 Dec 6 to 2018 Aug 14 UT, corresponding to $\delta t \approx 110.5-362.3$ days. The details of all of the {\it HST}/F606W observations are displayed in Table~\ref{tab:obs}. Results from these observations were previously reported in \citet{amb+18,max+18,lll+18,tvr+18,lll+19} and \citet{ptz+19}.

We processed all images in the same manner as described in \S\ref{sec:temp}. For observations taken within a few days of each other with the same instrument, we combine the visits to increase the signal-to-noise ratio and report them as a single epoch. In addition to our 2019 March observations, this also applies to observations over 2018 Jul 10-13 UT and 2018 Aug 14-15 UT.

\begin{figure*}[!t]
\includegraphics[width=\textwidth,trim={0in 6.25in 0in 0in}]{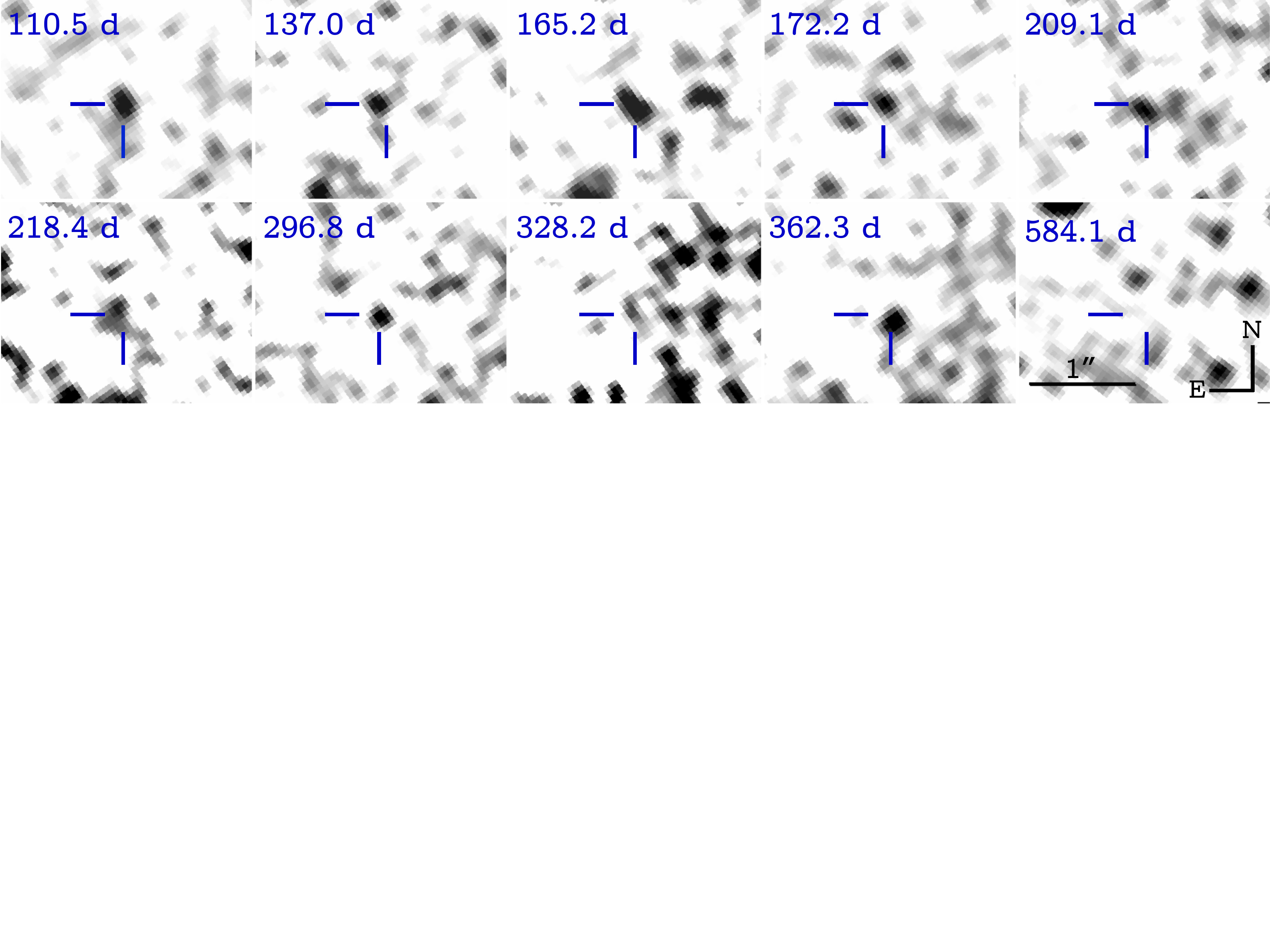} 
\caption{{\tt HOTPANTS} residual images from image subtraction between nine epochs of {\it HST}/ACS F606W imaging and the template observation obtained on 2019 Mar 21-27 UT (Program 15606). The last panel at $\delta t = 584.1$~days is the median-subtracted template. The position of the afterglow at $\delta t = 110.5$~days is denoted by the blue cross-hairs in all panels. The afterglow is detected at the $\gtrsim 3\sigma$ level in all residual images pictured here, while the template exhibits no source at the afterglow position to $m_{\rm F606W} \gtrsim 28.2$~mag. The scale and orientation of all images is denoted in the last panel, and all images have been smoothed with a 3-pixel Gaussian kernel.
\label{fig:image}}
\end{figure*}

\section{Analysis \& Results}
\label{sec:analysis}

\subsection{Astrometry}
\label{sec:astrometry}

We performed absolute astrometry of the first epoch in our sequence, 2017 December 6 UT, to the Pan-STARRS~1 catalog \citep{cmm+16} using 33 point sources in common with IRAF/{\tt ccmap} and {\tt ccsetwcs}. The resulting absolute astrometric tie uncertainty is $0.039''$ ($1\sigma$). We align all subsequent images to this epoch using the \texttt{tweakreg} task as part of the Drizzlepac package, which uses common sources to align the images in WCS to sub-pixel precision. For each image, {\tt tweakreg} uses 83-147 sources, with a relative astrometric tie error range of $4.8-11.0$~mas. Using Source Extractor \citep{SExtractor}, we derive an afterglow position based on the 2017 December 6 epoch of $\alpha=$\ra{13}{09}{48.07} and $\delta=-$\dec{23}{22}{53.37} (J2000) with an uncertainty of $0.040''$ (including a positional uncertainty for the afterglow centroid of $8.7$~mas). This position is consistent with that of the kilonova (e.g., \citealt{sha+17}) and we use it for our subsequent photometric analysis.

\subsection{A New Template}

To measure the upper limit on the afterglow at the position of \gw\ in the 2019 March observation, we first subtract off the smooth galaxy background which we model with a S\'{e}rsic surface brightness profile using the {\tt GALFIT} software package \citep{phi+07}. In {\tt GALFIT} we employ a point-spread function (PSF) empirically determined from stars in the image using IRAF/{\tt daophot}. We use IRAF/{\tt addstar} to inject artificial point sources at the position of \gw\ in the {\tt GALFIT} residual image, with the PSF determined above. We then perform photometry on each injected source using a $0.2''$ aperture and apply the appropriate aperture correction to correct to infinity \citep{sjb+05}. We repeat the experiment with sources of varying brightness to determine the flux level that would be recovered at the $3\sigma$ level, resulting in a $3\sigma$ upper limit of $m_{\rm F606W} \gtrsim 27.6$ at the position of \gw.

As previously discussed in \citet{bbf+17}, a S\'{e}rsic galaxy model provides an inadequate description of the galaxy light, which exhibits large-scale shell structure and dust lanes, apparent in the residuals and in the resulting goodness-of-fit value ($\chi^2_{\nu}=332$ for 4374194 d.o.f.). This structure, along with the small-scale brightness fluctuations, are known limitations for deriving a limit with the galaxy subtraction method.

Given the limitations of the simple analytic model of the galaxy light, we explore an alternative method to subtract the background to improve our limit. We apply a median filter to the original image using a $30 \times 30$~pixel box (corresponding to $1.5''$ or $\approx0.29$~kpc on a side) using IRAF/{\tt median}, where the box size is chosen so that no evidence of structure on the scale of the PSF is detectable in the median-filtered image. We then subtract the median-filtered image from the original image to produce a median-subtracted image suitable for photometry (Figure~\ref{fig:image}). 

We perform photometry of faint sources in the median-subtracted images using IRAF/{\tt phot}, finding a $3\sigma$ limit of $m_{\rm F606W} \gtrsim 28.2$~mag. We also inject fake point sources at the position of GW170817 using the empirically-derived PSF from the full image and recover a similar limit. To check the sensitivity of the result to the details of the filter, we also produce similar median-subtracted images made with 40 and 50-pixel filters. The final limit is not sensitive to these details, demonstrating that this method is robust. We use this limit for the remainder of our analysis, since the non-detection of any source to this limit makes this image suitable as a template.

\subsection{Image Subtraction and Afterglow Photometry}

We use the {\tt HOTPANTS} software package \citep{bec15} to subtract the 2019 March original image from each of the earlier epochs, and convolve each residual image to the pixel scale of the template ($0.05''$ pixel$^{-1}$). The residual images are shown in Figure~\ref{fig:image}. Although the majority of imaging was performed with the UVIS detector, and the template is taken with ACS, the difference in photometric calibration between ACS and UVIS is negligible in the F606W filter compared to the measured uncertainties in afterglow photometry (see below), with typical differences of $\lesssim\!0.04$~mag \citep{dm17}. Thus, we can reliably perform aperture photometry directly on the residual images.

Using IRAF/{\tt phot} package, we perform aperture photometry of the afterglow. We use a $0.3''$ aperture corresponding to 2.5$\times$FWHM, fixed at the position of the afterglow in all epochs. For each of the ACS epochs, we calculate aperture corrections by performing photometry in $0.3''$ and $0.5''$-radius apertures for 10-12 bright, unsaturated stars in each of the original fields, resulting in initial corrections of $\approx 0.01-0.03$~mag. We then apply tabulated encircled energy corrections to correct the $0.5''$ apertures to infinity \citep{boh16}. For UVIS, we use the tabulated corrections\footnote{http://www.stsci.edu/hst/wfc3/analysis/uvis\_ee} to correct the $0.3''$-radius apertures to infinity, typically $\approx 0.11-0.13$~mag.

\begin{figure}[!t]
\includegraphics[width=0.5\textwidth]{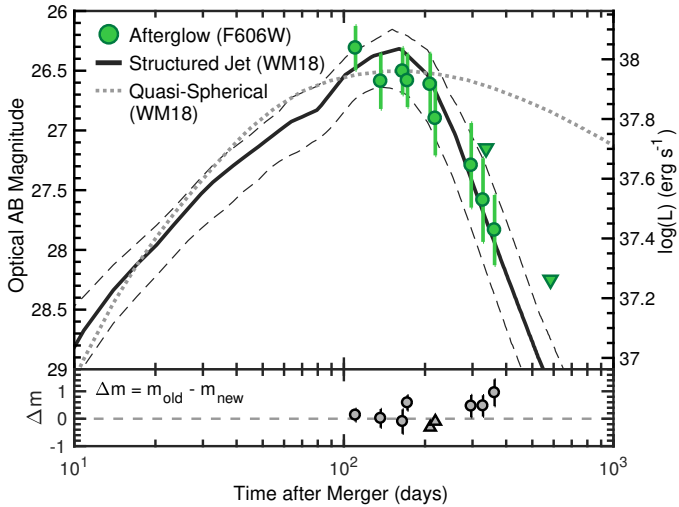}
\vspace{-0.1in}
\caption{{\it Top:} {\it HST}/F606W light curve of the afterglow of GW170817 spanning $\approx 110.5-584.1$~days (green points; observer frame); downwards triangles denote $3\sigma$ upper limits. The upper limit at $\approx584.1$~d is measured from the median-subtracted image, while all other data points are measured from {\tt HOTPANTS} residual images. Also shown are a structured jet model and the range of light curves describing the top $5\%$ of models (black solid and dot-dashed lines), and a quasi-spherical outflow model (dotted line; \citealt{wm18}). {\it Bottom:} Magnitude difference, $\Delta m$, between published values in previous works \citep{amb+18,max+18,lll+18,lll+19,ptz+19} and the new values measured in this work. Upward triangles denote epochs which were previously reported as upper limits, and are now detected in this work.
\label{fig:optlc}}
\end{figure}

In all except the epoch at $\approx 336.8$~days (2018 Jul 20), a source at the afterglow position is detected at the $\gtrsim 3\sigma$ level. The non-detection in that single epoch is unsurprising given the relatively shallow depth of the image (Table~\ref{tab:obs}). To derive the upper limit for this epoch, we performe aperture photometry of faint sources near the position of the afterglow.  The resulting photometry is listed in Table~\ref{tab:obs} and is displayed in Figure~\ref{fig:optlc}. For comparison, we also show the structured jet and quasi-spherical models that best fit the radio through X-ray evolution to $\approx 260$~days \citep{wm18}.

\begin{figure}[t]
\centering
\includegraphics[width=0.47\textwidth,trim={0in 0in 0in 0in}]{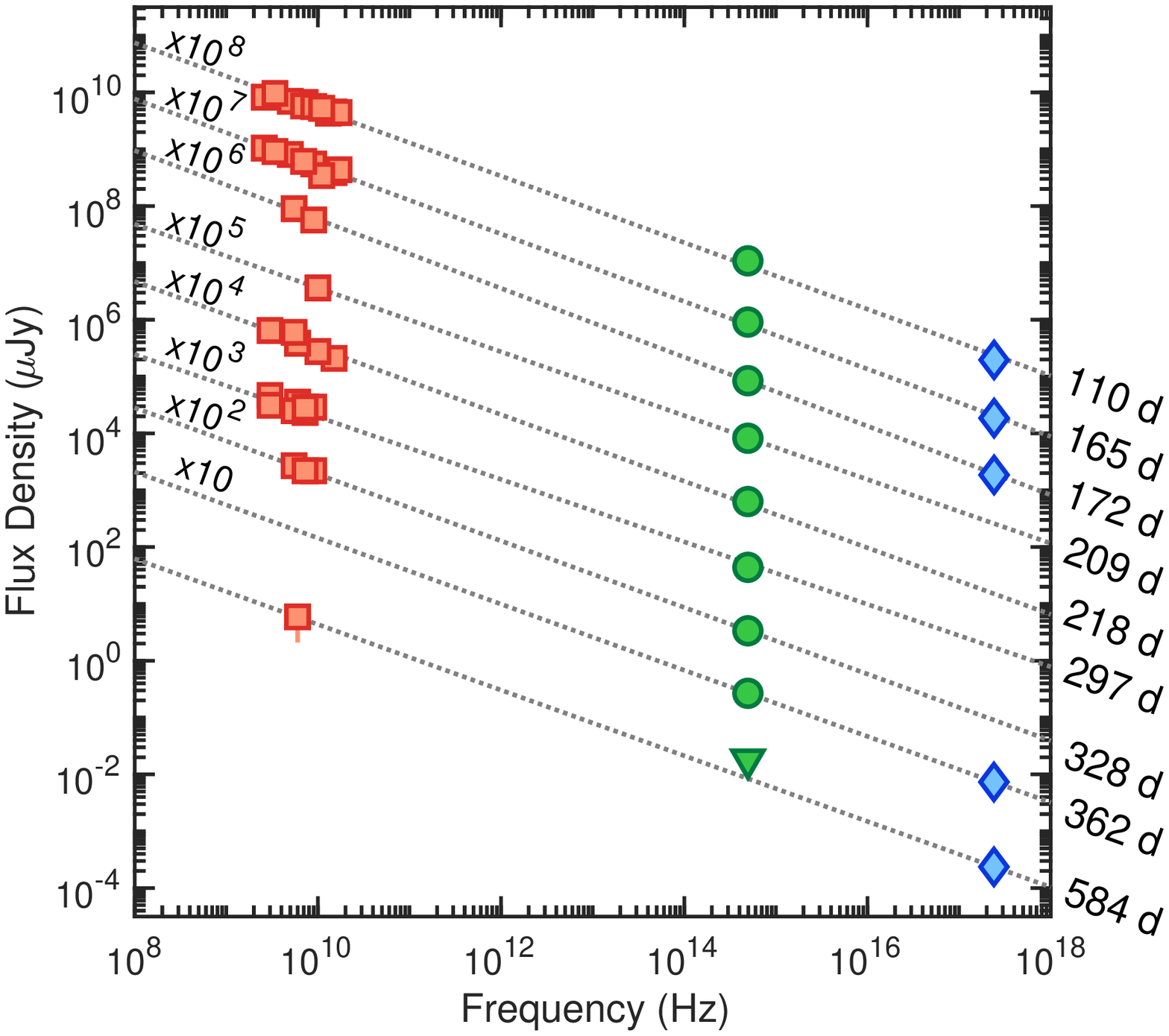} 
\vspace{-0.2in}
\caption{Broad-band SED of the afterglow of GW170817 at nine epochs of our {\it HST} observations, spanning $\approx 110-584$~days; fluxes are scaled for clarity. The {\it HST} photometry in this paper (green circles), radio afterglow (red squares; \citet{max+18,mnh+18,dkm+18,mfd+18,amb+18,tvr+18}, Hajela~et~al. in prep.), and X-ray afterglow (blue diamonds; Hajela~et~al. in prep.) are shown. The gray lines are best-fit power laws to the data at each epoch. $1\sigma$ uncertainties are plotted but the large majority are smaller than the size of the symbols.
\label{fig:broadband}}
\end{figure}

\subsection{Broad-band Afterglow Fitting}

To place the {\it HST} photometry in the context of the broad-band afterglow and quantify the broad-band spectral evolution at $\delta t \gtrsim 100$~days, we collect fluxes from the literature in the radio and X-ray bands at contemporaneous epochs, defined here to be within $\pm 10$~days of {\it HST} observations. In the radio band, there are available data for all epochs except at $\delta t \approx 137$, $337$~days, and $362$~days. The data are taken with the Karl G. Jansky Very Large Array (VLA) and the Australia Telescope Compact Array (ATCA), spanning 2.5-17 GHz \citep{amb+18,dkm+18,max+18,mdg+18,mfd+18,mnh+18,tvr+18}. We also use a 6~GHz VLA observation at $\delta t \approx 585$~days, presented in Hajela et al. (in prep.).

In the X-ray band, we find relevant comparison {\it Chandra X-ray Observatory} observations at five epochs. Previous analyses of these observations have appeared in \citet{nrh+18,max+18,tpr+18,pkw+18,rnh+18,tvr+18,lyk19}. Here, we use the fluxes and spectral parameters calculated in Hajela~et~al. (in prep.), which serves as a uniform analysis of all available {\it Chandra} data of the X-ray afterglow of \gw\ to $\approx 583.1$~days. To enable comparison of the X-ray observations to the optical and radio data, we convert the $0.3-10$~keV X-ray fluxes to flux densities, $F_{\nu,X}$, at a fiducial energy of 1~keV, using the derived photon index, $\Gamma$ at each epoch, where $F_{\nu,X} \propto \nu^{\beta_X}$ and $\beta_X \equiv 1-\Gamma$. The radio and X-ray data, along with our {\it HST} photometry, are displayed in Figure~\ref{fig:broadband}. 

We use $\chi^2$-minimization to fit the broad-band spectrum at each epoch to a single power law model in the form $F_{\nu} \propto \nu^{\beta}$, characterized by spectral index $\beta$ and a flux normalization parameter. We fit all of the available data at each epoch separately. The resulting fits have $\chi^2_{\nu} \approx 0.6-1.3$, demonstrating that the single power law model is adequate to fit the data over all epochs (Figure~\ref{fig:broadband}). The values for $\beta$ and $1\sigma$ uncertainties are given in Table~\ref{tab:beta} and the temporal evolution is displayed in Figure~\ref{fig:beta}. We calculate a weighted average of the spectral index across all epochs considered here of $\langle \beta \rangle = -0.583 \pm 0.013$.

\begin{deluxetable}{ccc}[t]
\tabletypesize{\normalsize}
\linespread{1.2}
\tablecolumns{3}
\tablewidth{0pc}
\tablecaption{Broad-band Spectral Index $\beta$
\label{tab:beta}}
\tablehead{
\colhead {$\delta t^\dagger$}	 &
\colhead {$\beta$}  &
\colhead {Data Reference$^\ddagger$} \\
\colhead {(d)}	 &
\colhead {} 		&	
\colhead {} 	
}
\startdata
110.49 & $-0.586_{-0.044}^{+0.024}$ & 1-3 \\
 165.19   & $-0.594_{-0.053}^{+0.032}$ & 1, 3 \\
  172.21   & $-0.606_{-0.032}^{+0.020}$ & 3, 4 \\
  209.10   & $-0.562_{-0.028}^{+0.020}$ & 5 \\
  218.37   & $-0.586_{-0.12}^{+0.053}$ & 5-7 \\
  296.80   & $-0.549_{-0.13}^{+0.057}$ & 5-6, 8  \\
  328.22   & $-0.586_{-0.097}^{+0.053}$ & 8 \\
  362.32   & $-0.582_{-0.057}^{+0.067}$ & 3 \\
  584.13  & $-0.578_{-0.040}^{+0.061}$ & 3 \\
  \hline
\multicolumn{1}{l}{All, weighted avg.} &  $-0.583 \pm 0.013$ \\
\enddata
\tablecomments{$^\dagger$ This is the epoch of the {\it HST} observation. Radio and X-ray observations with $\delta t \pm 10$~days were considered contemporaneous and were included in the power-law fits.  \\
$^\ddagger$ Literature references for the plotted radio and X-ray data. All {\it HST} data points are from this work. \\
{\bf References:} (1) \citet{max+18}; (2) \citet{mnh+18}; (3) Hajela et al. (in prep.); (4) \citet{dkm+18}; (5) \citet{mfd+18}; (6) \citet{amb+18}; (7) \citet{ptz+19}; (8) \citet{tvr+18}  }
\vspace{-0.4in}
\end{deluxetable} 

\section{Discussion}
\label{sec:disc}

\subsection{Off-Axis Afterglow Properties}
We present a revised light curve of the optical afterglow of \gw, relative to previous studies which have used subsets of {\it HST} observations to derive measurements and upper limits of the afterglow in the F606W filter \citep{amb+18,max+18,lll+18,lll+19,ptz+19}. We calculate the difference $\Delta m$ between the published values and the values presented in this work (Figure~\ref{fig:optlc}). Overall, we find that the afterglow in most epochs is systematically brighter than previously reported, with differences of $\Delta m \approx -0.1-1$~mag between published values and the values presented in this work (Figure~\ref{fig:optlc}), and an increase in $\Delta m$ as the afterglow becomes fainter. Our analysis also recovers a $\gtrsim 3\sigma$ source in two observations which were previously reported as upper limits \citep{amb+18,ptz+19}. The differences with respect to published values are not surprising given the non-uniformity of methods used for both galaxy subtraction and photometry, and the complicated structure of NGC4993 which makes accurate galaxy subtraction, and thus background estimation, challenging without a proper template.

The temporal evolution of the optical afterglow exhibits a flattening at $\approx 110-172$~days, followed by a steep decline at $\gtrsim 200$~days (Figure~\ref{fig:optlc}). The entire data set can be fit with a broken power-law with $\alpha_1 = -0.4 \pm 0.2$ and $\alpha_2 = -2.20 \pm 0.25$ (where $F_{\nu} \propto t^{\alpha}$), with a break time of $\approx 200-240$~days depending on the smoothness of the break. The large change in slope signifies that we are witnessing the peak, followed by a rapid decline after the jet break \citep{rho99}. For synchrotron emission, the post jet-break decline is expected to evolve as $F_{\nu} \propto t^{-p}$, where $p$ is the electron power-law index describing the input energy distribution of electrons \citep{sph99}, and thus we can infer a value of $p = 2.20 \pm 0.25$ from the optical light curve.

\begin{figure}[!t]
\centering
\includegraphics[width=0.45\textwidth,trim={0in 0in 0in 0in}]{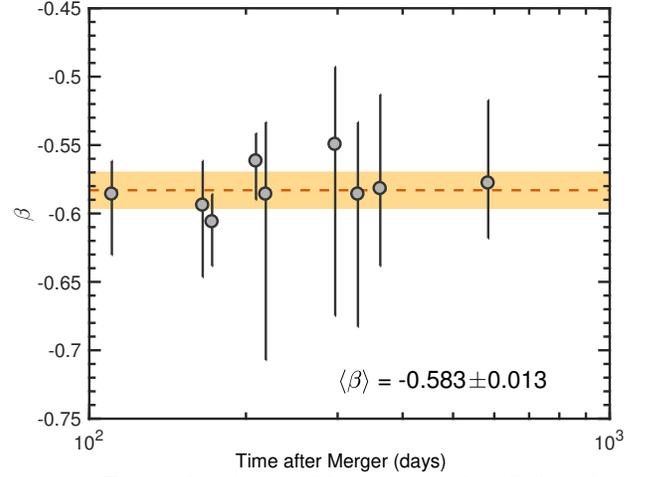} 
\vspace{-0.1in}
\caption{Temporal evolution of the spectral index, $\beta$, from fitting the radio, {\it HST} and {\it Chandra} X-ray data. Uncertainties correspond to $1\sigma$, and are produced from the $\chi^2$ fitting procedure. The red dashed line and orange band denotes the weighted average and uncertainty across the $\approx 110-584$-day interval.
\label{fig:beta}}
\end{figure}

We can obtain an independent constraint on $p$ from the spectral behavior of the source. Combined with the radio and X-ray evolution, the afterglow of \gw\ maintains the same spectral index within $1\sigma$ uncertainties for the duration of the {\it HST} observations (Figure~\ref{fig:broadband}-\ref{fig:beta}). This demonstrates that the radio, optical and X-ray bands all lie on the same spectral slope between $\nu_m \lesssim \nu \lesssim \nu_c$ (where $\nu_m$ is the peak frequency and $\nu_c$ is the cooling frequency of the synchrotron spectrum; e.g., \citealt{spn98,gs02}) out to $\approx 584$~days, and that no break frequencies evolve between the radio and X-ray bands on these timescales. The inferred value of $p=1-2\beta = 2.166 \pm 0.026$, is fully consistent with the value derived from the light curve, as well as with previous works based on broad-band data out to $\approx 260$~days \citep{max+18,amb+18,wm18,lll+19}.

A comparison of the {\it HST} light curve to models which best fit the radio and X-ray light curves to $\approx 260$~days \citep{max+18,wm18,amb+18,ktg+19} demonstrates that the optical emission at $\gtrsim 100$~days is emanating from a relativistic structured jet viewed off-axis (Figure~\ref{fig:optlc}). Moreover, models of mildly-relativistic quasi-spherical outflows, in which the jet (if produced at all) fails to break out of the ejecta, over-predict the observed optical flux by $\gtrsim 1.5-4$~times at $\gtrsim 200$~days. This provides clear confirmation from the optical emission that we are viewing an off-axis jet as opposed to a quasi-spherical outflow. This supports previous studies which reached a similar conclusion based primarily on temporal and spectral behavior within a single band \citep{dcs+18,lll+18,mfd+18,nrh+18,lll+19} and broad-band data \citep{amb+18,lpm+18,max+18,tvr+18,wm18}. This is also corroborated by the detection of super-luminal motion and constraints on the jet size from very long baseline interferometric observations \citep{mdg+18,gsp+19}, although it has also been shown that the source motion is expected to be indistinguishable between jetted and quasi-spherical models for $\lesssim 300$~ days \citep{zxm18}.

\begin{figure}[t]
\centering
\includegraphics[width=0.48\textwidth,trim={0in 0in 0in 0in}]{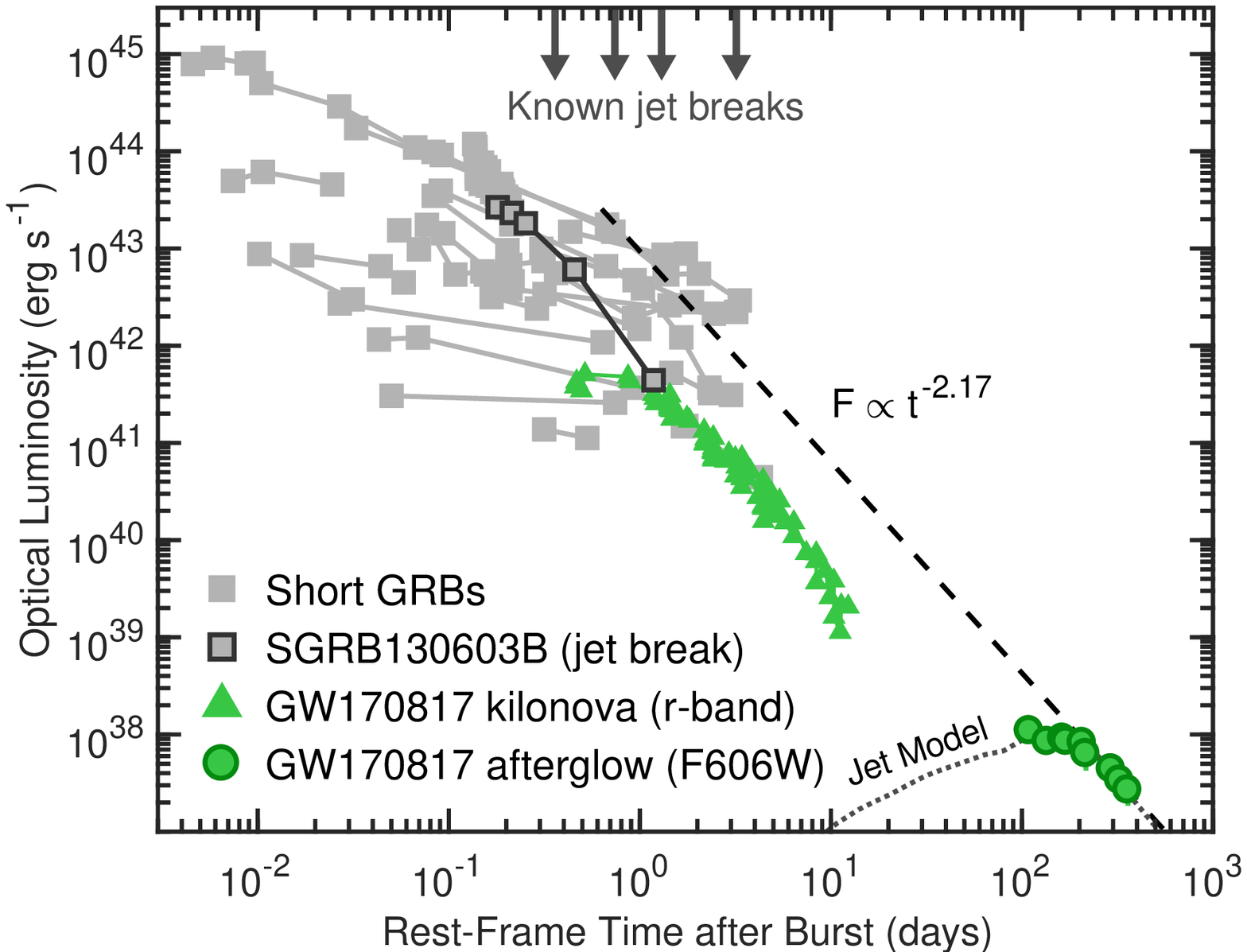} 
\vspace{-0.2in}
\caption{The afterglow (this work) and $r$-band kilonova (compiled in \citealt{vgb+17}, see references in text) of \gw\ along with the structured jet model (dotted line, \citealt{wm18}). Also shown are 25~short GRBs with optical afterglow light curves; GRB\,130603B is the single known jet break in the optical band and is highlighted. Arrows from the top denote the rest-frame jet break times of four short GRBs. The extrapolation of the post-peak slope of \gw\ ($\alpha = -2.17$) back to the luminosities of short GRBs intersects the population at $\approx 0.7-4$~days.
\label{fig:sgrb_lc}}
\end{figure}

\subsection{Comparison to Short GRB Afterglows}

\gw\ represents the first detection of an off-axis optical afterglow, while cosmological short GRBs represent those events seen close to or on-axis. The presence of relativistic jets in both types of events is one of several characteristics that signify a common origin. The similarity in their inferred explosion properties also suggest that the primary difference in behavior between short GRB jets and that of \gw\ is the viewing angle \citep{fbb+17,kbg18,lpm+18,sga+19,wm19}.

Here, we explore this in another way by connecting the optical properties of \gw\ with short GRBs. If \gw\ and cosmological short GRBs share the same values for their explosion properties, in particular a combination of the jet opening angle, kinetic energy and circumburst density, then regardless of observer angle, the post-jet break behavior of their afterglows should asymptote to the same declining light curve at late times \citep{vm12}. In this case, an extrapolation of the post-peak decline of \gw\ should intersect with the short GRB population at their expected jet break times in luminosity-space.

We collect data of all short GRBs with multiple optical afterglow detections, comprising 25~events (updated from \citealt{fbm+15}, and including the afterglow and kilonova of the short GRB\,160821B; \citealt{ltl+19,tcb+19}). We use the burst redshifts to obtain the afterglow luminosities as a function of rest-frame time, assuming $z=0.5$ (the median value of the population; \citealt{ber14,fbb+17}) for bursts without determined redshifts. The short GRB light curves are shown in Figure~\ref{fig:sgrb_lc}, highlighting the single source with a jet break measured in the optical band (GRB\,130603B; \citealt{fbm+14}). The extrapolation of the post-peak slope of $\alpha \approx -p \approx -2.17$ from \gw\ intersects short GRBs at $\approx 0.7-4$~days (rest-frame). Indeed, the short GRBs with measured jet breaks have a range of jet break times that are similar, $\approx 0.4-3.5$~days (Figure~\ref{fig:sgrb_lc}; \citealt{bgc+06,sbk+06,fbm+12,fbm+14,tsc+16}). This simple exercise is consistent with the notion that the combination of jet opening angle, kinetic energy, and circumburst density of \gw\ are similar to the population of short GRBs. As learned from short GRBs, we expect there to be inherent diversity in these properties that will manifest itself as a spread in behavior \citep{fbm+15}.

A further comparison of optical emission from short GRBs to the kilonova of \gw\ (compiled in \citealt{vgb+17}) clearly demonstrates that for on-axis events, short GRB afterglows are likely to outshine their optical kilonovae at all epochs if the luminosity and evolution of \gw\ is representative of the population. However, the overlap between the kilonova of \gw\ and the low-luminosity end of the short GRB distribution, including the claimed kilonovae in the short GRBs\,150101B and 160821B \citep{trp+18,ltl+19,tcb+19} leaves open the possibility that a small subset of short GRBs are discovered slightly off-axis and the optical emission is in fact dominated by the kilonova in these cases. Finally, Figure~\ref{fig:sgrb_lc} shows that if \gw\ had not been in solar conjunction at $\approx 15-100$~days, we would have been able to witness the rise of the optical afterglow starting at $\approx 20$~days and potentially the intersection with the kilonova emission.

\subsection{Constraints on a Globular Cluster Origin}

\begin{figure}[t]
\centering
\includegraphics[width=0.48\textwidth,trim={0in 0in 0in 0in}]{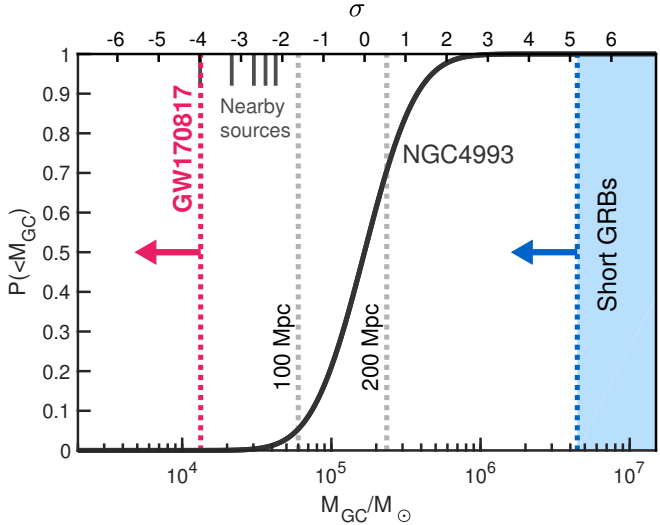} 
\vspace{-0.25in}
\caption{The GCMF of the host galaxy NGC4993 (black line), derived from the GCLF \citep{lki18}, compared to various limits: the limit at the position of \gw\ on a star cluster of $\lesssim 1.3 \times 10^{4}\,M_{\odot}$ (red dotted line), upper limits from $z\lesssim 0.3$ SGRBs (blue region), and the limits on GCs for similarly deep {\it HST} observations for events at 100~Mpc and 200~Mpc (dotted gray lines). Also shown are the corresponding masses of nearby objects (if they are GCs) at the distance of \gw\ (gray lines from top). The top axis denotes the $\sigma$ from the mean of the NGC4993 GCMF for a Gaussian distribution. Compared to the GCMF, the observations rule out a cluster at the position of \gw\ at a level of $\sim 4\sigma$ below the mean.
\label{fig:gcmf}}
\end{figure}

We now explore \gw\ in the context of its progenitor formation. Previous studies have used the stellar mass, stellar population age, star formation history of NGC4993, and the location of \gw\ with respect to the host galaxy center, to infer properties of the progenitor system, including the kick velocity, helium-star mass, and initial separation \citep{gw170817progenitor, bbf+17}. Overall, these studies found consistency between the progenitor properties and the distributions of Galactic binary neutron stars which formed via isolated binary evolution (e.g., \citealt{wwk10}). On the other hand, early simulations of interactions in globular clusters (GCs) have suggested that their dense stellar environments can also provide a significant channel of neutron star mergers and short GRBs, through dynamical encounters, tidal capture, or in-cluster primordial evolution (e.g., \citealt{gpm+06,ihr+08,lr10}). Moreover, the observed double neutron star system B2127+11C in the GC M15 is expected to merge within a Hubble time \citep{agk+90,tkf+17}, motivating a search for a GC at or near the position of \gw.

At the distance of \gw, GCs would appear unresolved or marginally resolved in our {\it HST} imaging depending on the S/N of the source. To compare the {\it HST} limit to the luminosities of GCs, we adopt the globular cluster luminosity function (GCLF) derived from prior ACS/F606W imaging. The GCLF is characterized by a Gaussian in magnitude space with a mean and width of $m_{\rm F606W}= 25.45 \pm 0.69$~mag \citep{lki18}. Using $M_{\odot,{\rm F606W}}=4.72$~mag \citep{wil18}, this translates to log($L/L_{\odot}$) = $4.92 \pm 0.27$. Adopting a mass-to-light ratio of $\approx 2 M_{\odot}/L_{\odot}$ \citep{ssl+09,bau17} the globular cluster mass function (GCMF) can be approximated as a Gaussian with a mean and width of log($M/M_{\odot}$) = $5.22 \pm 0.27$. We note that we do not carry out an independent GCLF determination based on our imaging as this would yield incremental returns compared to \citet{lki18}, due to the difficulty in confirming the GC nature of sources well below the GCLF peak.

With our deep {\it HST} observation at $\approx 584$~days, we place a constraint of $M_{\rm F606W} \gtrsim -4.8$~mag, or $L \lesssim 6.7 \times 10^{3}\,L_{\odot}$, on any underlying cluster. It is instructive to compare this limit to the GCMF as generally the rate of in-cluster interactions, and thus mergers, increases with cluster mass \citep{pla+03}. A comparison to the GCMF of NGC4993 places a limit of $M_{\rm GC} \lesssim 1.3 \times 10^{4}\,M_{\odot}$, $\approx 4\sigma$ below the mean; only $\approx0.004\%$ of the total mass in GCs in NGC4993 is below this limit (Figure~\ref{fig:gcmf}). This limit is also constraining enough to rule out $\approx\!70\%$ of the mass function of young massive clusters, corroborating the lack of any young stellar populations in the galaxy  \citep{ps10,bbf+17,llth+17}.

To place this limit in the context of previous limits from cosmological short GRBs, we search for the deepest available optical limits on persistent sources from low-redshift ($z \lesssim 0.3$) events obtained from previous optical imaging. We find that the most constraining limit is from GRB\,050709 at $z=0.161$ \citep{ffp+05}, which corresponds to $M\lesssim 4.5 \times 10^{6}\,M_{\odot}$, $\approx 5\sigma$ above the GCMF mean for NGC4993\footnote{We note that for the less massive star-forming host galaxy of GRB\,050709, the peak of its GCMF is expected to be similar, while the width may be narrower (e.g., \citealt{bs06}). In this case, the limit would correspond to $\gtrsim 5\sigma$ when compared to the GCMF of its host galaxy.} (Figure~\ref{fig:gcmf}). Thus, while short GRBs remain too distant to offer a firm conclusion on progenitor formation channels from direct imaging, the deep observations presented here place a direct and strong constraint on an {\it in situ} GC origin for a binary neutron star merger. While previous limits have been placed on an existing GC using more shallow, pre-explosion imaging \citep{bbf+17,llth+17,pks+17} as well as indirect inference from the fading behavior of the afterglow \citep{lll+19}, our analysis provides the deepest existing limit based on direct imaging of the event location.

Finally, we explore the possibility that the progenitor system of \gw\ was formed in a GC and ejected before merger (e.g., \citealt{bkl+14,am19}), using the median-subtracted 2019 March observation to identify nearby potential GCs. Using aperture photometry, we identify six sources with $\gtrsim 3\sigma$ significance within a projected distance of $\lesssim 400$~pc ($\lesssim 2.1''$). One of the sources clearly has an extended PSF (previously identified as a GC candidate in \citealt{pks+17}) and is most likely a background galaxy, while the five remaining sources are too faint to constrain their PSFs. If they are in fact GCs, their inferred masses are $\approx (1-4)\times 10^{4}\,M_{\odot}$ and contain at most $\approx 0.5\%$ of the GC mass of NGC4993, making it unlikely for the progenitor to have formed there.  In general, the progenitor system would have to travel at a minimum of the escape velocity of the GC, a few tens of~km~s$^{-1}$ for typical GC masses and sizes, and for a potentially long and uncertain merger timescale. Coupled with the old stellar population of the host galaxy, $\approx 11$~Gyr \citep{bbf+17}, it would thus be extremely challenging to correlate \gw\ with its parent GC. 

Looking forward, {\it HST} imaging to similarly deep limits of future well-localized gravitational wave events will provide meaningful limits on an {\it in situ} GC origin to 200~Mpc (assuming that the GCMF across galaxies is fairly constant; cf. \citealt{sbs+06}). Specifically, {\it HST} observations to $\approx 28.5$~mag for events at $\lesssim 100$~Mpc ($\lesssim 200$~Mpc) will be sensitive to $\gtrsim\!95\%$ ($\gtrsim 30\%$) of the GCMF (Figure~\ref{fig:gcmf}).

\section{Conclusions \& Future Outlook}
\label{sec:conc}

We present the first observation following \gw\ in which an optical source is not detected to deep limits, allowing us to determine the complete F606W light curve of its optical afterglow from $\approx 110-584$~days. The afterglow evolution is fully consistent with the optical emission emanating from a relativistic structured jet at an observer angle of $\approx 30^{\circ}$, as indicated by radio and X-ray observations. This study highlights the importance of template observations in determining accurate light curves, especially for the late and faintest stages of evolution. This is especially important for local events detected by gravitational wave facilities which are embedded in their host galaxies, for which galactic low surface brightness features are more prominent and cannot be easily modeled.

We also compare \gw\ to on-axis cosmological short GRBs. Extrapolating the optical post-peak temporal evolution of \gw\ to the luminosities of short GRBs, the predicted jet break times for short GRBs are consistent with their observed breaks. Thus, we find that the two populations can be easily connected if their explosion properties (e.g., energetics, circum-merger densities and jet opening angles) are similar, and that the factor which primarily dictates their different evolution is the observer angle. Continued studies of short GRBs to $\gtrsim 5$~days, as well as similarly in-depth studies of local binary neutron star mergers, will continue to shed light on any intrinsic differences in these populations.

We provide a deep and direct constraint on the presence of an underlying globular cluster to $M \lesssim 1.3 \times 10^{4}\,M_{\odot}$, providing direct evidence that \gw\ did not form and merge in a cluster {\it in situ} at the $4\sigma$ level. However, we cannot place meaningful constraints on the possibility that the progenitor system was dynamically formed and ejected from its parent cluster. Future simulations which calculate accurate rates of such systems taking into account the full cluster evolution, coupled with further observational constraints on mergers at $\lesssim 200$~Mpc, will help to elucidate this formation channel.

Finally, we remark that {\it HST} had a singular role in the optical afterglow of the relatively nearby \gw. As gravitational wave facilities increase in sensitivity, most binary neutron star mergers will be detected farther away. If the optical luminosity of the \gw\ afterglow is representative, the advent of extremely large telescopes and future space-based initiatives, such as {\it JWST} will play an incredibly important role in the detection and characterization of off-axis afterglows from binary neutron star (and neutron star-black hole) mergers.

\section*{Acknowledgments}
\noindent W.F. acknowledges support by the National Science Foundation under grant Nos. AST-1814782 and AST-1909358. Support for Program number 15606 was provided by the National Aeronautics and Space Administration through grant HST-GO-15606.001-A from the Space Telescope Science Institute, which is operated by the Association of Universities for Research in Astronomy, Incorporated, under NASA contract NAS5-26555, and Chandra Award Number G09-20058A issued by the Chandra X-ray Center, which is operated by the Smithsonian Astrophysical Observatory for and on behalf of NASA under contract NAS8-03060. Based on observations made with the NASA/ESA Hubble Space Telescope, obtained from the data archive at the Space Telescope Science Institute. The new {\it HST} data used in this paper can be found in MAST: \dataset[https://doi.org/10.17909/t9-6qez-fw41]{https://doi.org/10.17909/t9-6qez-fw41}. This research was supported in part by the National Science Foundation under Grant No. NSF PHY-1748958. The National Radio Astronomy Observatory is a facility of the National Science Foundation operated under cooperative agreement by Associated Universities, Inc. MN is supported by a Royal Astronomical Society Research Fellowship. Research by DJS is supported by NSF grants AST-1821987, AST-1821967, AST-1813708, AST-1813466, and AST-1908972. JS acknowledges support from the Packard Foundation. KDA acknowledges support provided by NASA through the NASA Hubble Fellowship grant HST-HF2-51403.001 awarded by the Space Telescope Science Institute, which is operated by the Association of Universities for Research in Astronomy, Inc., for NASA, under contract NAS5-26555. The Berger Time-Domain Group is supported in part by NSF grant AST-1714498 and NASA grant NNX15AE50G. The Margutti group at Northwestern acknowledges support provided by the National Aeronautics and Space Administration through grant HST-GO-15606.001-A, through Chandra Awards Number GO9-20058A, DD8-19101A and DDT-18096A issued by the Chandra X-ray Center, which is operated by the Smithsonian Astrophysical Observatory for and on behalf of the National Aeronautics Space Administration under contract NAS8-03060. R.C. acknowledges support from NASA Chandra grant GO9-20058B.

\vspace{5mm}
\facilities{HST (ACS, WFC3), MAST, VLA, ATCA, Chandra}

\software{
IRAF \citep{iraf1,iraf2}, Drizzlepac \citep{gon12}, GALFIT \citep{phi+07}, Source Extractor \citep{SExtractor}
          }

\end{document}